\documentclass[epj]{svjour}
\usepackage[english]{babel}
\usepackage{graphicx}
\usepackage{color}
\usepackage{multirow}

\begin{document}

\title{Branching ratio of the super-allowed $\beta$ decay of $^{10}$C} 

\author
{B.~Blank\inst{1} \and
M. Aouadi\inst{1} \and
P.~Ascher\inst{1} \and
M.~Gerbaux\inst{1} \and
J.~Giovinazzo\inst{1} \and
S.~Gr{\'e}vy\inst{1} \and
T.~Kurtukian Nieto\inst{1} \and
M.R. Dunlop\inst{2} \and
R. Dunlop\inst{2} \and
A.T. Laffoley\inst{2} \and
G.F. Grinyer\inst{3} \and
P. Finlay\inst{4}{\footnotemark}
}

\institute{Centre d'Etudes Nucl\'eaires de Bordeaux Gradignan,
UMR 5797 CNRS/IN2P3 - Universit\'e de Bordeaux, 19 Chemin du Solarium, CS 10120, F-33175 Gradignan Cedex, France  \and
Department of Physics, University of Guelph, Guelph, Ontario N1G 2W1, Canada \and
Department of Physics, Univeristy of Regina, Regina, SK S4S 0A2, Canada \and
Department of Physics and Astronomy, KU Leuven, Celestijnenlaan 200 D, B-3001 Leuven, Belgium }

\abstract{
In  an experiment performed at the ISOLDE facility of CERN, the super-allowed $\beta$-decay branching ratio of $^{10}$C was 
determined with a high-precision single-crystal germanium detector. In order to evaluate the contribution of the pile-up of 
two 511~keV $\gamma$ quanta to one of the $\gamma$-ray peaks of  interest at 1021.7~keV, data were not only taken with $^{10}$C, 
but also with a $^{19}$Ne beam. The final result for the super-allowed decay branch is 1.4638(50)\%, in agreement with the average
from literature.
}

\authorrunning{B. Blank {\it et al.}}
\titlerunning{Branching ratio of the super-allowed $\beta$ decay of $^{10}$C}

\maketitle
\renewcommand{\thefootnote}{\fnsymbol{footnote}}
\footnotetext{a) Present address: Xanadu, 777 Bay Street, Toronto, Ontario, M5G 2C8, Canada}

\section{Introduction}

Super-allowed 0$^+ \rightarrow$ 0$^+ \beta$ decay is a powerful tool to explore properties of the fundamental 
weak interaction. The\-se transitions have been used to test the conserved vector current (CVC) hypothesis as well as to determine
the vector coupling constant $G_v$ and the V$_{ud}$ Cabbibo-Kobayashi-Mas\-kawa quark mixing matrix element~\cite{hardy15}.
To achieve these goals, the $ft$ value of 0$^+\rightarrow$ 0$^+$ $\beta$ decays has to be measured precisely, which requires 
precise knowledge of the $\beta$-decay Q value, half-life and  super-allowed branching ratio for as many 
nuclei as possible. Once this is achieved, the corrected $\mathcal{F}$t values can be determined:
$$
    Ft=ft(1+\delta_R')(1+\delta_{NS}-\delta_C)=\frac{k}{2G_v^2(1+\Delta_R^v)}
$$
where $\delta_R', \delta_{NS}$, and $\Delta_R^v$ are radiative corrections and $\delta_C$ is an isospin breaking corrections.
With these corrected $\mathcal{F}$t values, physics beyond the standard model (SM) of particle physics can be explored.

These extensions of the SM can be of different nature, one of them being the addition of scalar or tensor currents to the 
well-known vector and axial-vector currents. The possible addition of a small scalar contribution to the Fermi transitions can 
be tested with 0$^+ \rightarrow$ 0$^+ \beta$ decay. With only the vector current contributing, the corrected $\mathcal{F}$t values
should be constant, whereas the addition of a scalar term yields $\mathcal{F}$t values which, due to an additional term
in the Fermi function, are dependent on the Q value of the decay. This can be observed in particular for nuclei with small
Q values, i.e., in the series of well-known 0$^+ \rightarrow$ 0$^+ \beta$ decays, for the lightest super-allowed emitters $^{10}$C and 
$^{14}$O (see Figure 7 in ~\cite{hardy15}).

\begin{figure}[hht]
\begin{centering}
\includegraphics[scale=0.8,angle=0]{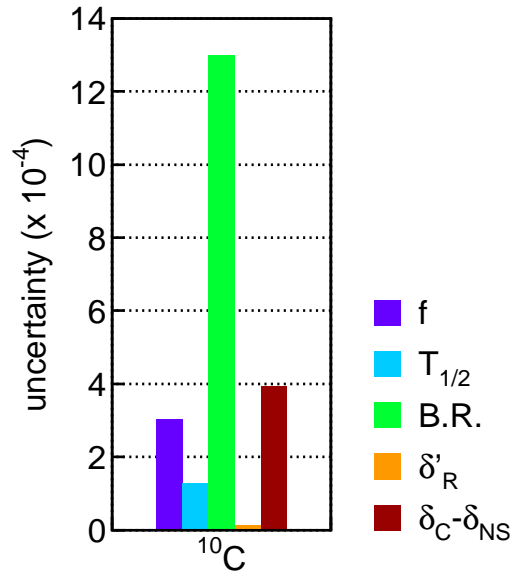}
\par\end{centering}
\caption{{\small{}}Error budget for the $\mathcal{F}$t value of $^{10}$C. By far the highest contribution comes from 
                   the super-allowed branching ratio of $^{10}$C.}
\label{fig:errors}         
\end{figure}

\begin{table*}[htt]
\caption{{\small{}\label{tab:c10_br}}
         Literature values for the super-allowed branching ratio of $^{10}$C prior to the present experiment.}
\begin{center}
\begin{tabular}{cccccccc}
\hline 
\hline 
\rule{0pt}{1.3em}
Sherr    & Freeman  &  Robinson &  Kroupa & Nagai & Fujikawa & Savard & average \\
et al. \cite{sherr53}& et al. \cite{freeman69} &  et al. \cite{robinson72} &  et al. \cite{kroupa91}&
et al. \cite{nagai91}& et al. \cite{fujikawa99}&  et al. \cite{savard95} &  \cite{hardy15}\\
 1.65(20) \%                  & 1.523(30) \%                     &    1.465(14) \%                    &   1.465(9) \% &
 1.473(7) \%                  & 1.4625(25) \%                    &    1.4665(38) \%               & 1.4646(19) \%\\
\hline 
\hline 
\end{tabular}
\end{center}
\end{table*}

Before the present work, the world data for $^{10}$C decay were as follows~\cite{hardy15}:
\begin{itemize}
\item the Q value is Q$_{EC}$~= 1907.994(67) MeV yielding a statistical rate function of f~= 2.30169(70)
\item the half-life is T$_{1/2}$~= 19301.5(25)~ms~\cite{dunlop16}
\item the super-allowed branching ratio is BR~= 1.4646(19)~\%. This value stems effectively from two experiments 
performed with large germanium detector arrays. Values of 1.4665(38)\%~\cite{fujikawa99} and 1.4625(25)\%~\cite{savard95} were obtained. 
Table~\ref{tab:c10_br} gives all literature values.
\item the theoretical corrections are  $\delta_R'$~= 1.679(4)\% and $\delta_C - \delta_{NS}$~= 0.520(39)\%.
\end{itemize}

The value of $\Delta_R^v$ is presently discussed in several theoretical papers~\cite{seng18,czarnecki19,gorchtein19,seng19,seng20}
and has a considerable uncertainty. However, as it is not affecting the determination of the corrected $\mathcal{F}$t value
for $^{10}$C, we refrain from commenting on its value.

\begin{figure}[hht]
\begin{centering}
\includegraphics[scale=0.5,angle=-90]{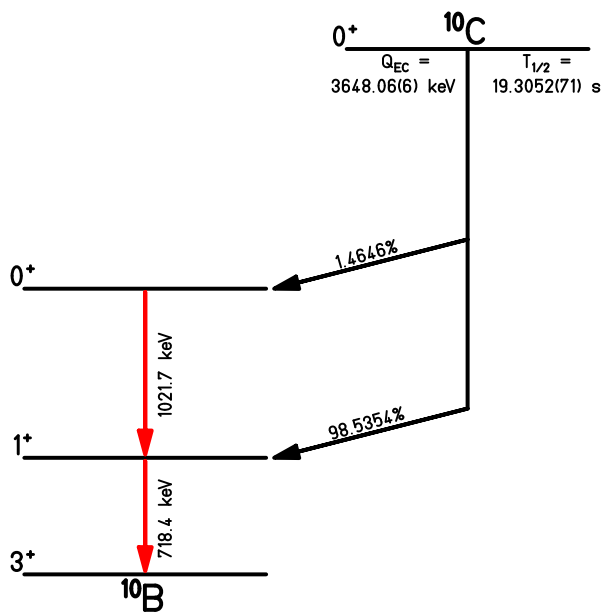}
\par\end{centering}
\caption{{\small{}}Partial decay scheme of $^{10}$C. The quantities measured in the present experiment are given in red.}
\label{fig:c10}         
\end{figure}

In Figure~\ref{fig:errors}, we plot the uncertainties of these different inputs for the determination of the $\mathcal{F}$t
value of $^{10}$C. The branching ratio is by far the most important contributor to the error bar of the $\mathcal{F}$t value
for this nucleus. The aim of the present work is to contribute to the improvement of the super-allowed branching ratio of
$^{10}$C. Although the decay scheme of this nucleus is quite simple (see Figure~\ref{fig:c10}), the difficulty in the present 
endeavour arises from the fact that the $\gamma$ ray at 1021.7~keV has to be measured in the presence of possible pile-up of two 
511~keV photons in the $\gamma$-ray detector. Therefore, in addition to the measurement of the $^{10}$C decay itself, additional 
measurements with $^{19}$Ne have been performed, a nucleus which is also a $\beta^+$ emitter with a half-life (17.22~s) and 
a $Q$ value (3238.4~keV) similar to those of $^{10}$C but no $\gamma$ ray close to the pile-up energy.

\section{Experimental set-up}

The experiment was conducted at the ISOLDE facility of CERN. A 1.4~GeV proton beam from the PS-Booster impinged 
on a CaO target with an maximum intensity of about 2$\mu$A for the production of $^{10}$C and about a factor of 
5 less for the production of $^{19}$Ne. $^{10}$C and $^{19}$Ne were ionised with a VD7 plasma ion source~\cite{penescu10}. 
After mass selection with the ISOLDE high-resolution separator HRS, the beam was sent to the LA1 
experimental station, where the detector set-up was mounted. The ISOLDE beam gate was constantly 
open to accept the full intensity extracted from the ISOLDE target. However, the number of proton pulses in a CERN
"supercycle" was varied over the whole experiment ranging from 3 out of about 30 cycles per supercycle to half of
the cycles in a supercycle. For runs where we took half of the cycles in a supercycle (in fact one out of two cycles), 
the detection rate was constant over the course of the run. For runs with only a few cycles per supercyle, they were spread roughly regularly
over the supercycle (e.g. cycles 1, 11, and 21 for 30 cycles per supercycle) to have a detection rate as constant as possible
in our detectors.

There was no detectable contamination for the $^{19}$Ne beam, whereas the situation was worse 
for the $^{10}$C runs. At the beginning of the experiment, tests were made by the target team to check whether the production
of atomic $^{10}$C is more favourable than the production of $^{10}$C$^{16}$O molecules. It turned out that CO molecules are 
produced with an intensity about a factor 2 larger than atomic $^{10}$C. In addition, the HRS could not be set on masses 
as small as A=10. However, at mass 26, not only $^{10}$C$^{16}$O arrived at the detection station, but also 
$^{12}$C$^{14}$O and $^{13}$N$_2$. The production of $^{12}$C$^{14}$O was negligibly small (2.8$\times$10$^{-4}$ compared
to $^{10}$C$^{16}$O). $^{13}$N$_2$ was produced as much as $^{10}$C$^{16}$O and gave thus twice as much 511~keV
$\gamma$ rays.

\begin{figure}[hht]
\begin{centering}
\includegraphics[scale=2.5,angle=0]{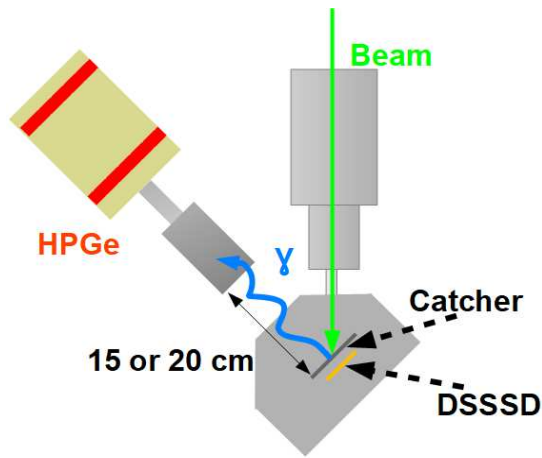}
\par\end{centering}
\caption{{\small{}}The experimental set-up of the present experiment. In the vacuum chamber, a catcher foil intercepted the $^{10}$C$^{16}$O and 
                   $^{19}$Ne beams. The implantation distribution was followed on-line by detecting the $\beta$ particles 
                   in a double-sided silicon strip detector installed about 1~mm behind the catcher foil. 
                   The precisely calibrated germanium detector was installed at a distance
                   of 15.00(1)~cm or 20.00(1)~cm from the catcher foil.}
\label{fig:setup}         
\end{figure}

The detection set-up (Figure~\ref{fig:setup}) consisted of a vacuum chamber with an aluminium catcher foil (200$\mu$m thickness) 
and a double-sided silicon strip detector (DSSSD, 500$\mu$m thickness) with 16 X and 16 Y strips and a pitch of 3mm, 
installed about 1~mm behind the catcher foil. The DSSSD served to optimise and control the implantation profile of 
the beam on the catcher foil. Gamma rays were detected outside the vacuum chamber (1.9~mm window of aluminium) 
by a precisely efficiency calibrated germanium detector~\cite{blank14ge} at a distance of 15 or 20 cm. 
The full-energy detection efficiency for the $\gamma$ rays of interest at 718.3~keV and 1021.7~keV were
0.28279(17)\% and 0.22348(15)\% at a distance of 15 cm as well as 0.17671(66)\%  and 0.14157(83)\% at 20 cm.

The germanium detector was precisely calibrated in efficiency at a distance of 15~cm~\cite{blank14ge}. If the detector model we use in the
simulations were perfect, the efficiencies calculated at a distance of 20~cm should be correct, too. However, an inspection of the
measurements at 15~cm and at 20~cm (see below) evidenced a systematically higher branching ratio at 20~cm. We therefore embarked
in a new series of calibration measurements of the germanium detector efficiency at 20~cm. For this purpose, we performed measurements
at 15~cm and 20~cm with sources of $^{60}$Co, $^{137}$Cs, and $^{207}$Bi. The measurements with the two $^{60}$Co lines and the
1063~keV line of $^{207}$Bi allowed us to determine the efficiency ratio between the 15~cm and 20~cm positions for an energy close to
the 1022~keV line of $^{10}$C, whereas we used the $^{137}$Cs $\gamma$ ray at 662~keV and the $^{207}$Bi line at 570~keV to determine
the experimental efficiency ratio for the 718~keV $\gamma$ ray for $^{10}$C. We assumed that the variation of the ratios with energy
is sufficiently small so that the slightly lower calibration energies for the 718~keV line and the slightly higher calibration lines
for the 1022~keV line yield acceptable ratios for the energies of $^{10}$C.

The finding was that the calculated efficiencies of the 718~keV and 1022~keV lines at 20~cm were factors of 0.8(29)$\times$10$^{-3}$ and 
4.8(15)$\times$10$^{-3}$ too small, respectively. We therefore corrected the calculated
efficiencies at 20~cm with these factors. For the uncertainties, we added in quadrature the error bar
of the correction factor and the correction factor itself to the uncertainty of the calculated efficiency, which yields in the end
the larger error bars for the efficiencies at 20~cm.

Two data acquisitions (DAQ) were run in parallel. The first DAQ had a single parameter which was the germanium energy.
In addition, a scaler module was read for each event. This scaler counted the number of proton pulse sent to ISOLDE since the beginning 
of the run, the number of $\gamma$-ray triggers from the germanium detector and the time of each event with a precision of 
1~millisecond. This data acquisition was triggered only by the germanium detector.

The second data acquisition was only used on-line. It registered the germanium energy signal and the energy signals from the
32 strips of the DSSSD. Similar to the first DAQ, a scaler registered the proton pulses, the germanium triggers and the time. 
This DAQ allowed us to optimise the beam implantation profile by detecting the $\beta$-decay position profile with the DSSSD
and to supervise it during the experiment. It was triggered by the detection of a $\beta$ particle in the DSSSD.
After optimisation, the beam spot was centred on the catcher foil and had a size of about 8~mm (FWHM) in X and Y, 
negligible compared to the distance of the source from the detector.

\section{Data taking}

As mentioned in the introduction, the main difficulty in the present experiment is to correctly evaluate the 511~keV~- 
511~keV pile-up which adds to the 1021.7~keV peak from the decay of $^{10}$C. For this purpose, the $^{10}$C activity transported 
to the detection set-up was largely varied by taking more or less proton pulses per supercycle which modifies the 
pile-up probability as a function of the 511~keV detection rate per second squared. In addition, we modified once during the
experiment the distance between the radioactive sample and the germanium detector entrance window from 15~cm to 20~cm
therefore varying both the total rate and the pile-up probability. This pile-up probability is also directly proportional
to the shaping time of the germanium signal. We therefore used two shaping times for the germanium signal of 2$\mu$s and 
1$\mu$s. Table~\ref{tab:settings} gives details about the settings used in the different runs of the experiment.

The idea of these changes in decay rate in the set-up, in distance and in shaping time is that in the end we have
to get the same branching ratio, independent of the settings. This procedure is meant to search for systematic errors in 
our measurements and will be tested below.

\begin{table}[htt]
\caption{{\small{}\label{tab:settings}}
         Settings used throughout the experiment. Germanium signal shaping times and germanium distance
         were modified several times during the experiment.}
\begin{center}
\begin{tabular}{cccc}
\hline 
\hline 
\rule{0pt}{1.3em}
isotope   & run number &  shaping time &  distance   \\
 $^{10}$C &  42 - 55   &    2 $\mu$s   &   15 cm \\
 $^{10}$C &  56 - 97   &    1 $\mu$s   &   15 cm \\
 $^{10}$C &  98 - 116  &    2 $\mu$s   &   15 cm \\
$^{19}$Ne & 118 - 127  &    2 $\mu$s   &   15 cm \\
$^{19}$Ne & 128 - 133  &    1 $\mu$s   &   15 cm \\
$^{19}$Ne & 134 - 144  &    2 $\mu$s   &   20 cm \\
 $^{10}$C & 145 - 173  &    2 $\mu$s   &   20 cm \\
\hline 
\hline 
\end{tabular}
\end{center}
\end{table}

However, these changes were not enough to quantitatively evaluate the pile-up probability. For this purpose, we spent an 
important part of the beam time on measurements with $^{19}$Ne (about 18~h as compared to 78~h for $^{10}$C). 
$^{19}$Ne has decay characteristics similar to
$^{10}$C, however, without having a $\gamma$ ray at 1022~keV. Therefore, counts above background in this region can only come
from pile-up of two 511~keV $\gamma$ rays. By measuring the 511~keV and the 1022~keV rate, we are able to determine
the pile-up probability on an absolute scale.

\begin{figure}[hht]
\begin{centering}
\includegraphics[scale=0.5,angle=0]{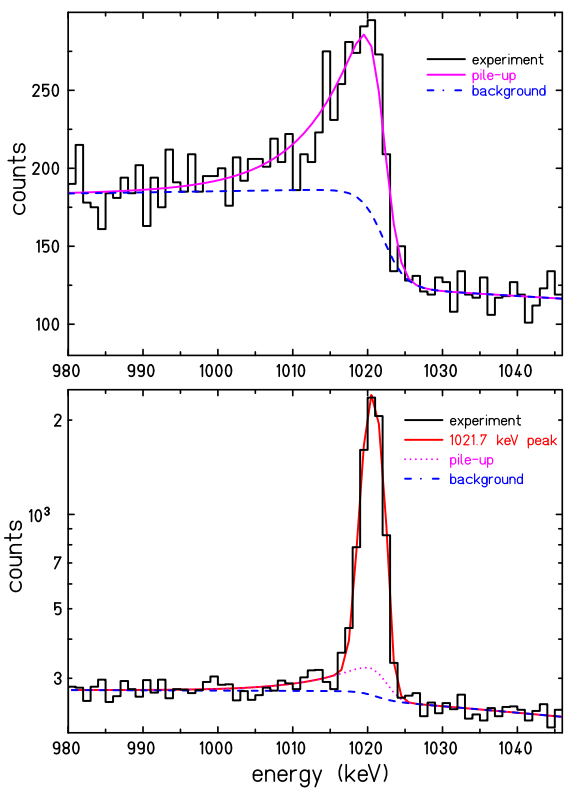}
\par\end{centering}
\caption{{\small{}}Gamma-ray spectrum in the region of the 1022~keV peak. The upper part shows a spectrum from 
                   a $^{19}$Ne run, where the pile-up contribution from two 511~keV $\gamma$ quanta and the background is 
                   visible. The lower spectrum contains in addition $\gamma$ rays at 1021.7~keV from the decay of the 
                   second excited state of $^{10}$C. The different curves give the different contributions to the spectrum. 
                   The pile-up contribution in the lower figure is calculated and subtracted before determining the integral
                   of the peak.}
\label{fig:1022}         
\end{figure}

Figure~\ref{fig:1022} shows the region around E$_\gamma$~= 1022~keV for a run with a $^{19}$Ne beam and with a $^{10}$C beam.
In the $^{19}$Ne case, the spectrum can be described by a background-contribution step function and a Gaussian with 
a low-energy tail. In the case of $^{10}$C, the additional peak at 1021.7~keV is seen on top of these two contributions.

The pile-up correction as introduced below depends sensitively on the counting rate of the 511~keV $\gamma$ rays. Therefore, a
run selection was performed where only runs with a constant 511~keV $\gamma$-ray rate were kept. Changes in this rate were
caused by modifications of the CERN supercycle or other problems with the PS-Booster or the ISOLDE front-end during a run.
These problems forced us to remove 33 runs out of 99 for the $^{10}$C measurements and 1 run (out of 26) for the $^{19}$Ne measurements 
totalling about 34\% and 3.5\% of the running time for each isotope, respectively.

\begin{figure}[hht]
\begin{centering}
\includegraphics[scale=0.37,angle=-90]{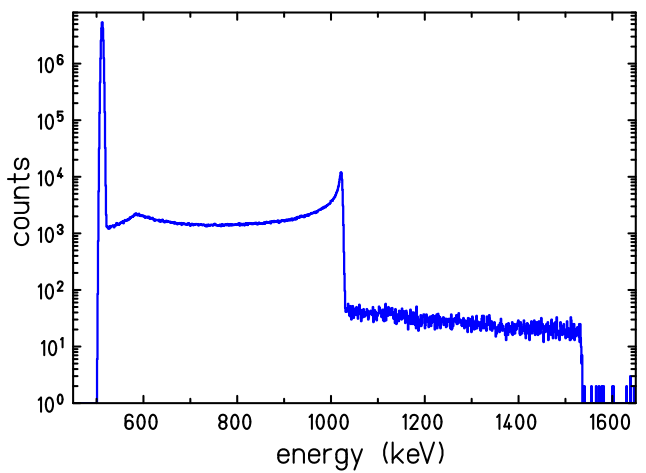}
\par\end{centering}
\caption{{\small{}}Simulated $\gamma$-ray spectrum for a single-energy $\gamma$ ray of 511~keV. The time distribution of the events
                   was taken from a $^{18}$Ne run (run 118, see figure~\ref{fig:cycles} for the time structure of this run). 
                   Depending on the time difference between two events, a total
                   signal energy between 1022~keV (no or negligible time difference between two events) and 511~keV (time difference longer
                   than the signal width) is found. Events above 1022~keV come from triple and quadruple coincidences. The edge at about
                   600~keV is due to the fact that we use a DAQ time window of 10$\mu$s. For a signal shaping time of 2$\mu$s,
                   the second event starts to be cut by this time constraint.}
\label{fig:pileup}         
\end{figure}

In order to correctly determine the number of counts under the 1021.7~keV peak, we have to subtract this pile-up contribution.
As mentioned earlier, the pile-up of two 511~keV $\gamma$ ray depends quadratically on the rate of these $\gamma$ rays 
and on the time difference between the two events. If the time difference is close to zero, an energy signal 
at 1022~keV results. The more the two $\gamma$ rays are separated in time, the lower is the signal energy with 
the lower limit being a 511~keV signal. This limit corresponds in fact to the case where there is no pile-up 
of the signals in the detector. The exact shape of the pile-up peak in the spectrum depends on the signal shape from
the electronics (in particular the shaping amplifier). 

Figure~\ref{fig:pileup} shows a simulation where only 511~keV $\gamma$ rays 
were considered. The events between 511~keV and 1022~keV come from two events which overlap more and more in time. At 1022~keV,
a perfect overlap is reached. Above 1022~keV, triple coincidences and above 1533~keV quadruple coincidences come into play.
The pile-up probability for each run can be determined by these simulations as the number of counts above the 511~keV
peak. It varied as a function of the total counting rate of the germanium detector between 0.2\% and 2\% for the different runs.
Details of the MC simulations will be given below.

\subsection{$^{19}$Ne runs}

\begin{figure}[hht]
\begin{centering}
\includegraphics[scale=0.37,angle=-90]{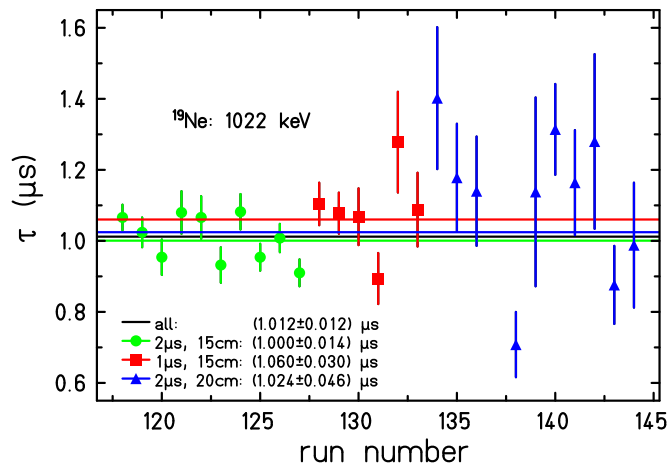}
\par\end{centering}
\caption{{\small{}}Pile-up times as determined from formula~\ref{eq:tau} for each $^{19}$Ne run. The average values for the 
                   different settings agree with each other satisfactorily, although the $\chi^2$ is only about 2 for the 
                   different averages. The values for the 1$\mu$s runs were multiplied by a factor 2.}
\label{fig:ne19}         
\end{figure}

We used the $^{19}$Ne runs to determine the pile-up probability. For this purpose, the pile-up peak at 1022~keV and the 511~keV
peak were fit to determine their functional form and their intensity. It was found that the shape of the 1022~keV
peak is always the same, independent of the signal shaping time or the pile-up rate. Only the pile-up probability, i.e. 
the number of counts in the pile-up peak, depends on these two parameters. Therefore, the 511~keV rate and the 1022~keV 
pile-up rate can be linked by the following formula:
\begin{equation}
N_{1022} = N_{511}^2 * \tau
\label{eq:tau}
\end{equation}
where N$_x$ are the rates of the 511~keV and 1022~keV peaks and $\tau$ is a pile-up time. We found in addition that $\tau$
is constant for a fixed experimental shaping time of the experiment electronics. To go from a shaping time of 2$\mu$s (most
of the runs) to 1$\mu$s, the pile-up time has simply to be divided by two. Figure~\ref{fig:ne19} shows the pile-up times
for all $^{19}$Ne runs as determined with the formula above. Although the $\chi^2$ of all the average values is about 2, we consider
that we can use a constant pile-up time for all runs.

\begin{figure}[hht]
\begin{centering}
\includegraphics[scale=0.31,angle=0]{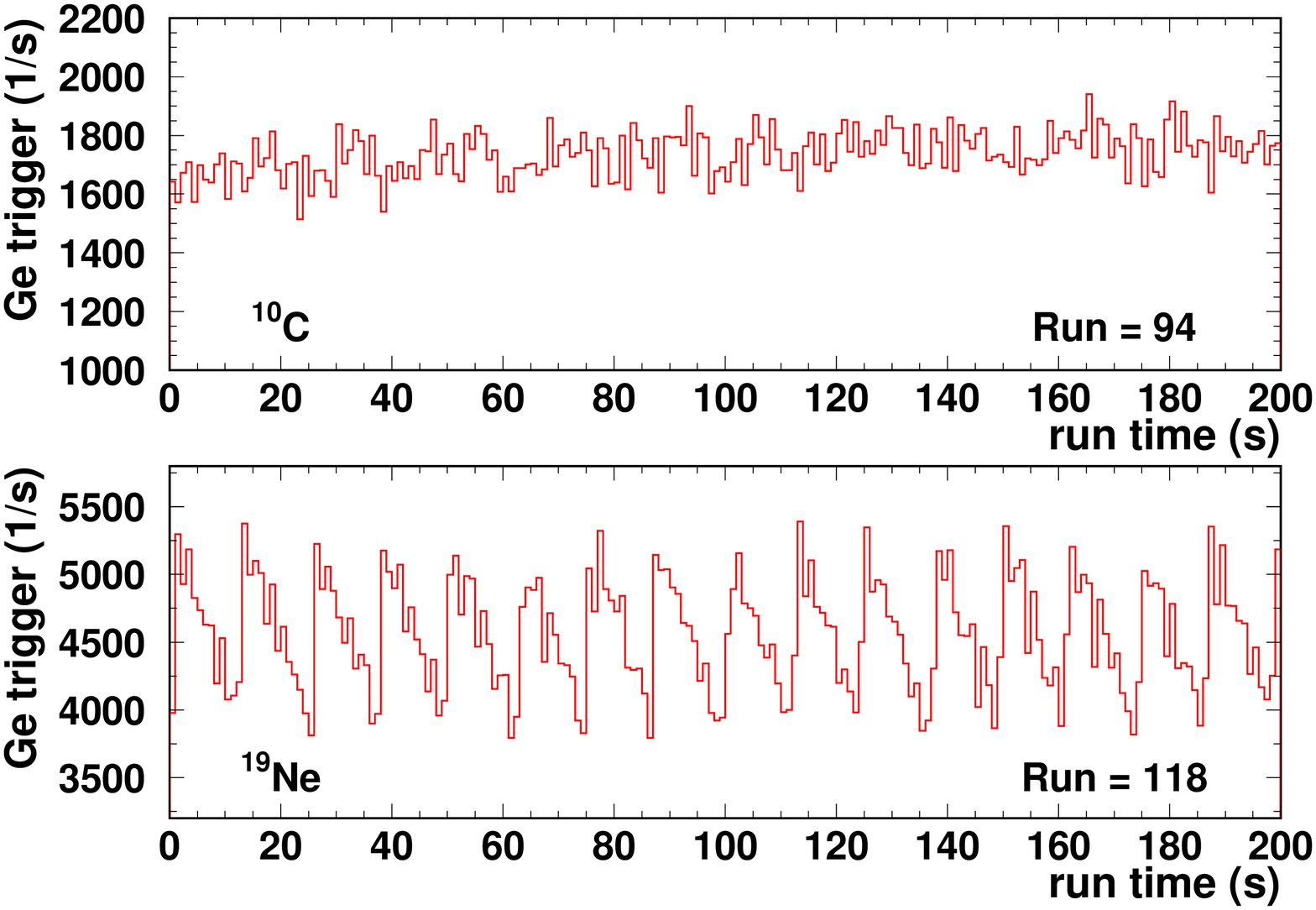}
\par\end{centering}
\caption{{\small{}}The two figures show the different event time structures as seen by the germanium detector. The upper part shows 
                   the germanium triggers as a function of time for a run with the $^{10}$C activity, whereas the lower spectrum shows 
                   the same information for a $^{19}$Ne run. Although both nuclei have approximately the same half-life, 
                   the time structure at the experiment is largely different due to the different release properties for CO molecules and
                   the associated contaminants
                   and $^{19}$Ne. In the latter spectrum, the impinging of the PS Booster proton beam on the ISOLDE target (every 12s and 13.2s,
                   multiples of the PS Booster time structure of 1.2~s) is clearly visible. For the $^{10}$C runs, the contaminants $^{14}$O and $^{13}$N with 
                   their longer half-lives also contribute to wash out the time structure. }
\label{fig:cycles}         
\end{figure}

Now that we have determined the exact functional form and the pile-up time with the $^{19}$Ne runs, we can use this information 
for the pile-up subtraction of the $^{10}$C runs (see Figure~\ref{fig:1022}). However, this procedure is only correct, 
if the time distribution is the same for both nuclei. An inspection of this time distribution (see Figure~\ref{fig:cycles})
shows that this is not at all the case. The noble gas character of $^{19}$Ne allows this isotope to diffuse and effuse in the
ISOLDE target - ion-source (TIS) ensemble extremely rapidly. Therefore, a well-pronounced time structure due to the proton impact
on the ISOLDE target and the $^{19}$Ne half-life is seen in the case of $^{19}$Ne (lower figure), which is completely absent in the 
case of the $^{10}$C$^{16}$O molecules and its contaminants, which are extracted from the TIS ensemble much slower that the proton beam time structure and the $^{10}$C
half-life.

As in the case of $^{19}$Ne this time distribution is not constant, a correction factor has to be applied when comparing $^{19}$Ne and
$^{10}$C. This factor will be determined in the following via Monte-Carlo (MC) simulations.

\subsection{Monte-Carlo simulations}

If the time structure of the $^{19}$Ne and the $^{10}$C runs were the same, the  pile-up probability determined with $^{19}$Ne could
be used directly for the $^{10}$C runs. However, this is not the case. Therefore, we developed a MC procedure to determine a 
correction factor to correlate the pile-up probabilities of $^{19}$Ne and $^{10}$C.

\begin{figure}[hht]
\begin{centering}
\includegraphics[scale=0.72,angle=0]{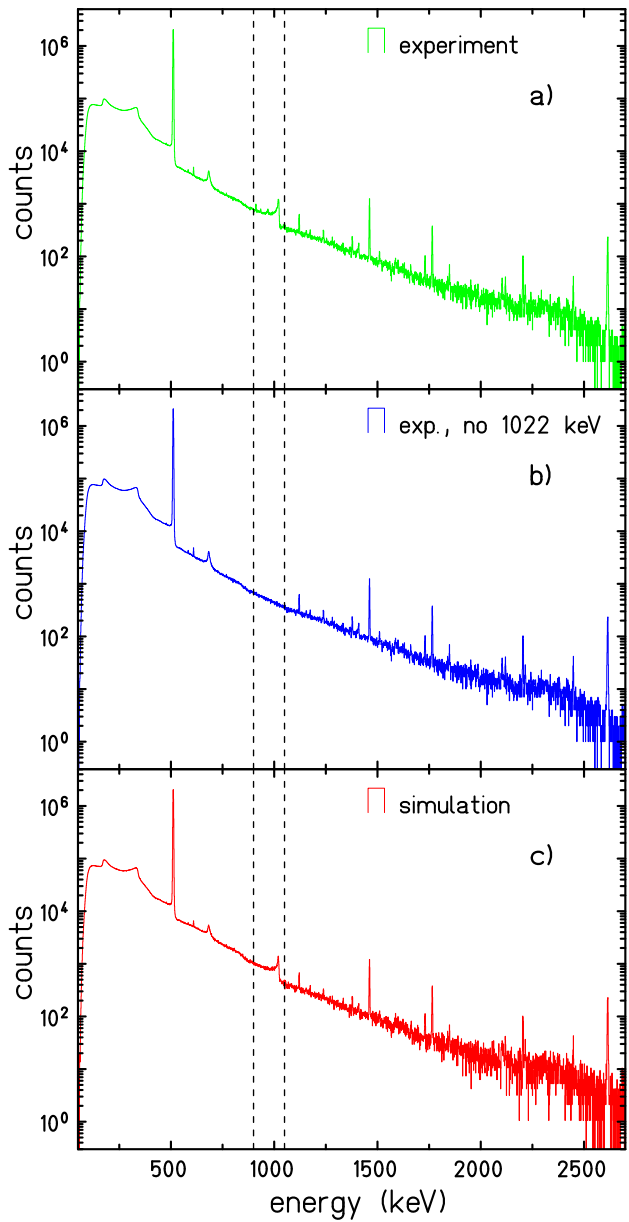}
\par\end{centering}
\caption{{\small{}}Experimental spectra and a spectrum generated by a Monte-Carlo procedure. a) Experimental spectrum from a $^{19}$Ne
                   run over basically the full energy range. b) Same spectrum, however, with the 1022~keV pile-up peak removed 
                   and the 511~keV peak increased to compensate for the removal of counts from this peak by pile-up
                   (see text for details). c) Simulated spectrum from a Monte-Carlo simulation which uses as inputs the energy
                   distribution from the spectrum in b) and the event time structure as displayed in fig.~\ref{fig:cycles}. 
                   The dashed vertical lines indicate the region where the effect of pile-up of two 511~keV $\gamma$ rays was removed.}
\label{fig:exp_sim}         
\end{figure}

This MC procedure uses the energy spectrum of each individual run and the associated time structure we measured with
the scaler module with a precision of 1~ms. Within one millisecond, we distributed the events randomly. In order to simulate
the pile-up, we first removed from the experimental spectrum (see Figure~\ref{fig:exp_sim}a) all counts above background in the
region of the 1022~keV peak. For this purpose, we first set all channels between 900~keV and 1050~keV to zero and added then
a smooth background with statistical fluctuations. In addition, we increased the number of counts in the 511~keV peak by the 
pile-up probability (0.2\% - 2\% for the different runs, see figure~\ref{fig:pileup}, which allowed us to determine the percentage of 511~keV events
removed from the 511~keV peak due to pile-up) to produce a 511~keV peak "without the effect of pile-up losses in this peak". 
The result is shown in Figure~\ref{fig:exp_sim}b.
This is then the energy input for the MC simulations. When used together with the event time distribution we can generate
a simulated spectrum with its pile-up contribution (Figure~\ref{fig:exp_sim}c). 

The only free parameter in these simulations is the "simulation shaping time", a parameter in the functional form of the
germanium detector signal used in the simulations which is equivalent to the experimental shaping time and varies the time width of the signal.
It was adjusted such that it matched the time structure of the experimental signal. 

This procedure can be applied to the $^{19}$Ne
runs but also to the $^{10}$C runs, where in addition to the pile-up counts also the 1021.7~keV counts are removed. As we can not
remove all pile-up events from the input energy spectrum (we correct only the 511~keV and the 1022~keV regions), the simulated
spectrum differs from the experimental one by several aspects. For example, the simulated spectrum is generally shifted towards higher
energies, because the experimental input spectrum contains $\gamma$ rays that are already piled-up. The simulation further piles-up these 
same events and shifts events to higher energies. This is particularly visible below 511~keV where counts in the spectrum
are lost. However, as we are only interested in the regions around the 511~keV and the 1022~keV peaks, we believe
that this procedure is sufficiently correct.

\begin{figure}[hht]
\begin{centering}
\includegraphics[scale=0.37,angle=-90]{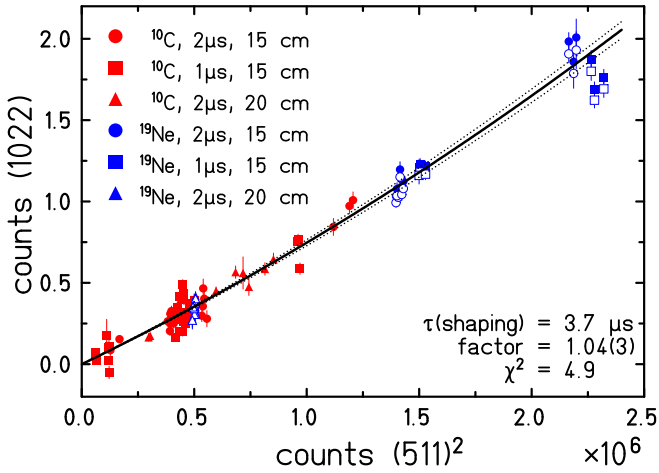}
\par\end{centering}
\caption{{\small{}}Results from the MC simulations of all runs (see text for details). The number of counts in the 1022~keV 
                   pile-up peak is plotted versus the number of counts in the 511~keV peak squared. This plot allows us to 
                   readjust the data from the $^{19}$Ne runs by a small factor to take into account the different event time
                   structures of the $^{10}$C and the $^{19}$Ne runs. The $^{19}$Ne data have to be increased by a factor 
                   of 1.04(3). The line is the fit of the data after correction with its uncertainty (dashed lines). The open blue symbols
                   are equivalent to the full blue symbols before multiplying them with the scale factor.}
\label{fig:pile_up}         
\end{figure}

Once these simulations were completed, the 511~keV and the 1022~keV peaks were fit as for the experimental spectra.
The integrals of these peaks are plotted in Figure~\ref{fig:pile_up}. In order to determine the matching factor between the
$^{19}$Ne and the $^{10}$C data, we fit these data with a second-order polynomial with a free scaling factor for the $^{19}$Ne data
to account for the different time structure.
The result is a scale factor of 1.04(3). The error bar was determined by varying the simulation shaping factor by $\pm$ 200~ns
which is roughly the parameter range where the experimental and the simulation signal shape are in good agreement.

\subsection{Pile-up test with the 718~keV $\gamma$ ray of $^{10}$C}

A direct test of the pile-up correction during the $^{10}$C runs is not possible with the 511~keV~- 511~keV pile-up, because on top of
the pile-up peak, the $\gamma$ rays of energy 1021.7~keV add. However, the pile-up can at least be checked to some extent with
the pile-up of two $\gamma$ rays of 718~keV. For these $\gamma$ rays, the statistics is about a factor of 10 lower than for 511~keV 
$\gamma$ rays, but it allows us nonetheless to test the correction.

\begin{figure}[hht]
\begin{centering}
\includegraphics[scale=0.39,angle=-90]{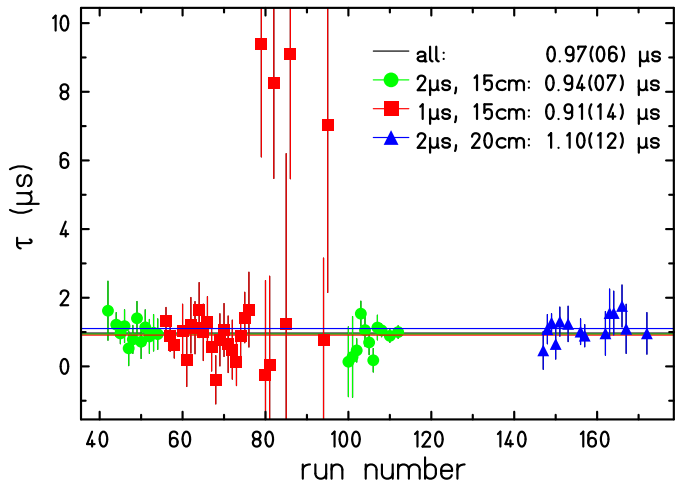}
\par\end{centering}
\caption{{\small{}}Pile-up times as determined for all $^{10}$C runs from the pile-up of two 718~keV $\gamma$ rays from the decay of the
                   first excited state of $^{10}$C. The runs from 78 to 98 are very low statistics runs and thus have large fluctuations.
                   The values for the 1$\mu$s runs were multiplied by a factor of 2.}
\label{fig:1436}         
\end{figure}

The same analysis as for the 511~keV pile-up has therefore been done also for this $\gamma$ ray. By means of equation~\ref{eq:tau}
we determined the pile-up time $\tau$ run by run. The result is shown in Figure~\ref{fig:1436}. The pile-up time is in perfect 
agreement with the value obtained with the 511~keV pile-up during the $^{19}$Ne runs and of the order of 1~$\mu$s (to be compared to
Figure~\ref{fig:ne19}).

\section{Results}

\subsection{Results with standard analysis parameters}

The results for the super-allowed branching ratio of $^{10}$C are obtained run by run as follows. For each $^{10}$C run, three peaks are
fitted: (i) the 511~keV annihilation peak, (ii) the 718~keV peak from the $\gamma$ decay of the first excited state of $^{10}$B
to its ground state, and (iii) the 1021.7~keV peak from the decay of the second excited state of $^{10}$B to its first excited state
after subtraction of the pile-up contribution. The pile-up contribution is obtained from the counting rate of the 
511~keV $\gamma$ rays by means of equation~\ref{eq:tau} and an additional small correction factor to take into account the  
difference in time structure between a $^{19}$Ne and a $^{10}$C run (see section 3.2). The number of pile-up counts thus determined is subtracted from the 
$\gamma$-ray spectrum with the correct pile-up peak shape after which the 1021.7~keV peak can be fitted with a Gaussian and a
background function.

\begin{figure}[hht]
\begin{centering}
\includegraphics[scale=0.36,angle=-90]{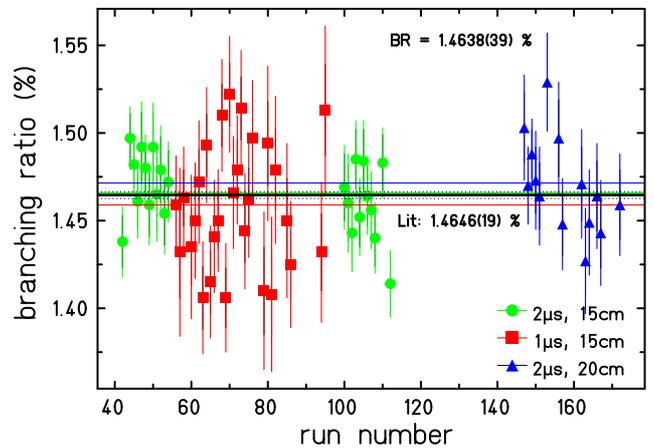}
\par\end{centering}
\caption{{\small{}}Super-allowed branching ratio as determined run by run. The average of all data is BR = 1.4638(39)\% with 
                   $\chi^2_\nu$~= 1.18 (already included in the error bar given). 
                   The individual averages are 1.4660(48)\% (2$\mu$s, 15 cm), 1.4591(65)\% (1$\mu$s, 15 cm),
                   and 1.4717(111)\% (2$\mu$s, 20 cm). The $\chi^2_\nu$ values are 0.99, 1.39, and 1.11 and were used to inflate the uncertainties quoted.}
\label{fig:br}         
\end{figure}

As there is no $\beta$-decay feeding of the ground state of $^{10}$B within the limits of the precision of our experiment, 
the super-allowed branching ratio of interest is obtained from the efficiency-corrected counting-rate ratio of 1021.7~keV 
to 718~keV $\gamma$ rays. Figure~\ref{fig:br} shows this branching ratio for the different $^{10}$C runs.
Averaging these run-by-run results yields a branching ratio of BR = 1.4638(39)\% with a normalised
$\chi^2_\nu$~= 1.18. This result contains already the increase of the uncertainty by the square root of the $\chi^2_\nu$.

\vspace*{-0.3cm}

\subsection{Search for systematic uncertainties}

The measurements with different electronics shaping times and with different distances between the activity and the germanium
detector agree with one another. The averaging of all individual runs yields a reduced $\chi^2_\nu$ of 1.18, 
which is a sign of good agreement of the branching ratios determined for the different runs. In addition, the branching ratios
determined as the averages for the different settings (two different shaping times, one additional position) agree with each other
(1.4660(48)\% (2$\mu$s, 15 cm), 1.4591(65)\% (1$\mu$s, 15 cm), and 1.4717(111)\% (2$\mu$s, 20 cm)).
Therefore, there is no reason to add a systematic uncertainty from the different experimental settings.

A first systematic uncertainty comes from the factor applied between the $^{19}$Ne and $^{10}$C pile-up correction.
As stated above, this factor is 1.04(3). By refitting the 1022~keV peak with factors of 1.01 and 1.07, we obtain
branching ratios of BR = 1.4664(40)\% and 1.4607(41)\%. This enables us to determine a systematic deviation with respect to 
the central value of $\pm$0.0029\%.

An additional systematic uncertainty comes from the shape of the pile-up peak that we determined with the $^{19}$Ne data.
It is determined by five parameters: (i) the position of the peak kept fixed once optimised, 
(ii) the sigma of the Gaussian kept also fixed because strongly correlated with the parameters of the tail function
(see next parameters), (iii) the step function for the background defined as a certain percentage of the peak height, 
(iv) the position of the tail on the low-energy side of the peak, and (v) the pile-up time ($\tau$ = 1.012(12)$\mu$s, see equation \ref{eq:tau}), 
which determines the number of pile-up counts from the number of counts from the 511~keV counting rate.
However, we do not consider the systematic error due to this last parameter, because this parameter being determined
in the fit of the $^{19}$Ne data the error is largely included in the variation of the $^{19}$Ne/$^{10}$C correction factor.
The systematic errors obtained by varying these parameters within their error bars are shown in table~\ref{tab:errors}.
With these systematic errors included, we obtain our final experimental result for the super-allowed branching ratio of $^{10}$C
of BR~= 1.4638(50)\%.

\begin{table}[htt]
\caption{{\small{}\label{tab:errors}}
         Error budget for the present experiment. The statistical uncertainty is summed quadratically with the
         different systematic uncertainties to yield a total uncertainty of 0.0050\%.}
\begin{center}
\begin{tabular}{cc}
\hline 
\hline 
\rule{0pt}{1.3em}
error type                   & uncertainty (\%)    \\
experimental statistics      & 0.0038    \\
efficiency simulation        & 0.0009    \\
pile-up correction: & \\
$^{19}$Ne/$^{10}$C pile-up factor    & 0.0029    \\
background step function     & 0.0001    \\
tail position                & 0.0010    \\
\hline 
total uncertainty            & 0.0050    \\
\hline 
\hline 
\end{tabular}
\end{center}
\end{table}

\section{Discussion and outlook}

Our new value of the super-allowed branching ratio of $^{10}$C is in agreement with the literature average of 1.4646(19)~\%. 
The error bar is a factor 1.5 to 2 larger than the most precise literature values~\cite{fujikawa99,savard95}. 
If we average our value with literature values with an uncertainty less than a factor of 10 larger than the smallest error bar, 
as prescribed in Ref.~\cite{hardy15}, we obtain a new average branching ratio of BR~= 1.4644(18)\%. 

Our result slightly modifies the previous world average. As our result is mainly statistics limited, 
a new experiment with an improved beam production and increased purity could allow to yield a competitive result.
As an improvement of the production rate of $^{10}$C seems to be difficult to achieve, a higher statistics experiment
will therefore require a longer beam time. However, as the present experiment already lasted 5.5 days, a factor of 2 more
beam time is certainly a limit. The measurement at 20~cm contributed little to the final result and should certainly be omitted
for a future experiment. It would be probably most reasonable to perform a future measurement only with a shaping time of 1~$\mu$s
which allows the highest rate with the smallest pile-up contribution. 

The production rate of $^{13}$N$_2$ was as strong as the rate of $^{10}$C-$^{16}$O. This yields twice as much 511~keV $\gamma$ rays from
$^{13}$N$_2$ than from $^{10}$C-$^{16}$O. The contamination from $^{13}$N$_2$ could be removed by the use of e.g. a Multi-reflection
Time-of-flight Spectrometer (MR-ToF-MS) or a Penning trap with a resolving power in excess of 10$^5$. If the 511~keV rate can thus be divided by a factor
of 3, the pile-up would decrease by a factor of 10 and become much less of a problem.

\section{Conclusion}

We have performed a measurement of the super-allowed $\beta$-decay branching ratio of $^{10}$C. After production and separation by the
ISOLDE facility, the $^{10}$C-$^{16}$O activity was constantly implanted in a catcher foil in front of our precisely efficiency 
calibrated germanium detector to measure the $\gamma$ rays emitted in the decay of $^{10}$C.

Our result of BR~= 1.4638(50)\% is a factor of 1.5 to 2 less precise than the most precise literature values. Nevertheless the 
present result demonstrates that the problems with the pile-up of two 511~keV $\gamma$ rays can be overcome. In a future 
experiment, a higher precision can be reached with a longer beam time and a better beam purification scheme. However, 
it will be very challenging to reach the same precision for the branching ratio as for the half-life and the Q value of $^{10}$C.

\section*{Acknowledgment}

We would like to thank the ISOLDE staff for their efforts during the present experiment. 
The research leading to these results has received funding from the European Union's Horizon 2020 research and innovation 
programme under grant agreement no. 654002.

\end{document}